\documentclass[a4,12pt]{article}
\usepackage{amsmath}

\begin{document}

\renewcommand{\theequation}{\arabic{section}.\arabic{equation}}

\title{Cosmological Models Generalising Robertson-Walker Models}
\author{Abdussattar\\
         Department of Mathematics\\ 
         Banaras Hindu University \\ 
         Varanasi - 221 005, India}
\date{}

\maketitle

\begin{abstract}
Considering the physical 3-space $t$ = constant of the space-time metrics as spheroidal and pseudo spheroidal, cosmological models which are generalizations of Robertson-Walker models are obtained. Specific forms of these general models as solutions of Einstein's field equations are also discussed in the radiation and the matter dominated era of the universe.
\end{abstract}


\section{Introduction}

In general relativity, a gravitational situation is equivalent to imposing a Riemannian structure on the space-time fourfold. There exists a correspondence between matter-energy and the geometry of the physical 3-space $t$ = constant. The space-times associated with the Robertson-Walker metrics, representing standard models of the universe, contain matter-energy to curve up the physical 3-space $t$=constant into 3-spheres and 3-pseudo spheres \cite{Nar93}. The absence of the dissipative terms in the energy-momentum tensor of the standard models, due to the assumption that the matter in the universe is homogeneous and isotropic, suggests that these models may not appropriately describe the early hot dense state of the big bang cosmology \cite{Wein72}. Cosmological models described by the space-times which have hypersurfaces $t$=constant as 3-spheroids and 3-pseudo spheroids are obtained in order to incorporate the dissipative terms in the energy-momentum tensor. Robertson-Walker models are special cases of these general models.

\section{Derivation of the metrics}

A 3-spheroid, immersed in a 4-dimensional Euclidean flat space given by the metric

\begin{equation}
  \label{eq:S2-1}
  (dS_E)^2 = (dx^1)^2 + (dx^2)^2 + (dx^3)^2 + (dx^4)^2
\end{equation}
in Cartesian coordinates $(x^1, x^2, x^3, x^4)$, will have the Cartesian equation
\begin{equation}
  \label{eq:S2-2}
  \frac{(x^1)^2 + (x^2)^2 + (x^3)^2}{R^2} + \frac{(x^4)^2}{S^2} = 1
\end{equation}
where $R$ and $S$ are constants. We use coordinates intrinsic to the surface of the 3-spheroid
\begin{eqnarray}
  \label{eq:S2-3}
 x^1 &=& R \sin\chi \cos\theta,\nonumber\\
 x^2 &=& R \sin\chi \sin\theta \cos\phi, \nonumber\\
 x^3 &=& R \sin\chi \sin\theta \sin\phi,\\
 \textrm{and }x^4 &=& S \cos\chi.\nonumber
\end{eqnarray}
The spatial line-element on the surface of the 3-spheroid is obtained from the metric (\ref{eq:S2-1}) of the 4-dimensional flat space in which it is immersed as
\begin{equation}
  \label{eq:S2-4}
  (dS_E)^2 = R^2 \left[\left\{\cos^2\chi + \frac{S^2}{R^2} \sin^2\chi \right\} d\chi^2 + 
             \sin^2\chi (d\theta^2+\sin^2\theta d\phi^2)\right]
\end{equation}
We now write $\sin\chi = r (0 \le r \le 1)$ and obtain $(dS_E)^2$ in $r, \theta, \phi$ coordinates as
\begin{equation}
  \label{eq:S2-5}
  (dS_E)^2 = R^2 \left[\left\{ 1+\frac{S^2}{R^2} \frac{r^2}{1-r^2}\right\} dr^2 +
             r^2 (d\theta^2 + \sin^2 \theta d\phi^2)\right]
\end{equation}
The space-time metric, which has its associated physical 3-space $t$=constant as a 3-spheroid, is therefore written as
\begin{equation}
  \label{eq:S2-6}
  ds^2 = -dt^2 + R^2 \left[\left\{ 1+\frac{S^2}{R^2} \frac{r^2}{1-r^2} \right\} 
                 dr^2 + r^2 (d\theta^2+\sin^2\theta d\phi^2) \right]
\end{equation}
where $R$ and $S$ are now functions of time $t$. Relativistic units are being used with $c = G = 1$.

Next we consider a 3-pseudo spheroid, with the Cartesian equation
\begin{equation}
  \label{eq:S2-7}
  \frac{(x^1)^2 + (x^2)^2 + (x^3)^2}{R^2} - \frac{(x^4)^2}{S^2} = -1
\end{equation}
where $R$ and $S$ are constants, immersed in a 4-dimensional pseudo Euclidean flat space with metric
\begin{equation}
  \label{eq:S2-8}
  (dS_p)^2 = (dx^1)^2 + (dx^2)^2 + (dx^3)^2 - (dx^4)^2
\end{equation}
We use coordinates intrinsic to the surface of the 3-pseudo spheroid
\begin{eqnarray}
  \label{eq:S2-9}
   .x^1 &=& R \sin h\chi \cos\theta,\nonumber\\
 x^2 &=& R \sin h \chi \sin\theta \cos\phi,\nonumber\\
 x^3 &=& R \sin h \chi \sin\theta \sin\phi,\\
\textrm { and } x^4 &=& S \cos h \chi.\nonumber
  \end{eqnarray}
The spatial line-element on the surface of the 3-pseudo spheroid is obtained from the metric (\ref{eq:S2-8}) of the 4-dimensional pseudo Euclidean flat space in which it is immersed as
\begin{equation}
  \label{eq:S2-10}
    (dS_p)^2 = R^2 \Bigg[\left\{ \cos h^2 \chi - \frac{S^2}{R^2}
               \sin h^2 \chi \right\} d\chi^2 +
 \sin h^2\chi (d\theta^2 + \sin^2 \theta d \phi^2) \Bigg]
\end{equation}
We now write $\sin h\chi = r$ and obtain $(dS_p)^2$ in $r, \theta, \phi$ coordinates as
\begin{equation}
  \label{eq:S2-11}
  (dS_p)^2 = R^2 \left[\left\{1-\frac{S^2}{R^2} \frac{r^2}{1+r^2} \right\}
             dr^2 + r^2 (d\theta^2 + \sin^2\theta d\phi^2)\right]
\end{equation}
The space-time metric, which has its associated physical 3-space $t$=constant as a 3-pseudo spheroid, is therefore written as
\begin{equation}
  \label{eq:S2-12}
  ds^2 = - dt^2+R^2 \left[\left\{ 1-\frac{S^2}{R^2} \frac{r^2}{1+r^2}\right\}
         dr^2 + r^2 (d\theta^2 + \sin^2\theta d\phi^2) \right]
\end{equation}
where $R$ and $S$ are now functions of time $t$. The two line-elements (\ref{eq:S2-6}) and (\ref{eq:S2-12}) can however, be combined into a single line-element with the help of a parameter $k$ that takes values $k = \pm 1$ as
\begin{equation}
  \label{eq:S2-13}
  ds^2 = -dt^2 + R^2 \left[\left\{1+\frac{S^2}{R^2} \frac{kr^2}{(1-kr^2)}\right\}
        dr^2 + r^2 (d\theta^2 + \sin^2 \theta d\phi^2)\right].
\end{equation}
It may be noted that if the parameter $k$ is set equal to zero we get the line-element (\ref{eq:S2-13}) as
\begin{equation}
  \label{eq:S2-14}
  ds^2 = -dt^2 + R^2 [dr^2 + r^2 (d\theta^2 + \sin^2\theta d\phi^2)]
\end{equation}
which has its associated physical 3-space $t$ = constant as a 3-surface of zero curvature. The space-time metrics (\ref{eq:S2-13}) in normal Gaussian coordinates $r, \theta, \phi, t$ are the natural generalizations of the Robertson-Walker space-times for which $S(t)=R(t)$. The signature of the space-time line-element (\ref{eq:S2-13}) is $(+,+,+,-)$ as $(1 + \frac{S^2}{R^2} \frac{kr^2}{(1-kr^2)})$ is always positive.

\section{The Field Equations}

\setcounter{equation}{0}

The energy-momentum tensor of a relativistic imperfect fluid is given by \cite{Wein72}
\begin{equation}
  \label{eq:S3-1}
  T^{ij} = pg^{ij} + (\rho+p)u^i u^j - \eta h^{ik} h^{jl} w_{kl} - \chi(h^{ik} u^j + h^{jk}u^i)
           q_k - \xi h^{ij} u^k; k
\end{equation}
where $w_{ij}$ is the shear tensor
\begin{equation}
  \label{eq:S3-2}
  w_{ij} \equiv u_{i;j} + u_{j;i} - \frac{2}{3} g_{ij} u^k;k,
\end{equation}
$q_i$ is the heat flow vector
\begin{equation}
  \label{eq:S3-3}
  q_i \equiv T_{;i} + T u_{i;j} u^j,
\end{equation}
$h_{ij}$ is the projection tensor on the hypersurface normal to $u^i$
\begin{equation}
  \label{eq:S3-4}
  h_{ij} \equiv g_{ij} + u_iu_j,
\end{equation}
$\chi$ is the coefficient of heat conduction, $\eta$ is the coefficient of shear viscosity, $\xi$ is the coefficient of bulk viscosity, $T$ is the temperature, $p$ is the isotropic fluid pressure, $\rho$ is the proper matter-energy density and $u^i$ is the fluid flow vector satisfying $g_{ij} u^i u^j = -1$. A semi-colon stands for the covariant differentiation. The thermodynamic behaviour of a dissipative fluid is governed by the following set of thermodynamic laws \cite{Wein72}
\begin{enumerate}
\item Particle number conservation:
  \begin{equation}
    \label{eq:S3-5}
    (nu^i)_{;i} = 0
  \end{equation}
\item The second law of thermodynamics:
  \begin{equation}
    \label{eq:S3-6}
    KTd\sigma = pd\left(\frac{1}{n}\right) + d\left(\frac{\rho}{n}\right)
  \end{equation}
\end{enumerate}
where $\sigma K$ is the entropy per particle. $K$ is the Boltzmann constant introduced here to make $\sigma$ dimensionless. An entropy currnet four-vector $S^i$ is defined by the equation
\begin{equation}
  \label{eq:S3-7}
  S^i = n K \sigma u^i - \chi T^{-1} q^i.
\end{equation}
The rate of entropy production per unit volume is given by $S^i_{;i}$. It may be noted that $S^i_{;i}$ is non-negative if $\chi \ge 0$. The models are also required to satisfy various physical conditions such as $0 \le p \le \frac{\rho}{3}$ and $\eta \ge 0, \xi \ge 0$.

In the adopted comoving coordinate system we have the components of the fluid 4-velocity vector $u^i$ as $u^1 = u^2 = u^3 = 0, u^4 = 1$ and $u_{i;j} u^j = 0$. The volume expansion $\theta \equiv u^i_{;i}$ for the models (\ref{eq:S2-13}) is obtained as
  \begin{equation}
    \label{eq:S3-8}
    \theta = \frac{3\dot{R}}{R} + \left(1 + \frac{S^2}{R^2} \frac{kr^2}{(1-kr^2)}\right)^{-1} 
             \left(\frac{\dot{S}}{S}
              - \frac{\dot{R}}{R}\right) \frac{S^2}{R^2} \frac{kr^2} {(1-kr^2)}
  \end{equation}
where an overhead dot denotes differentiation with respect to time $t$. The non-zero components of the shear tensor $w_{ij}$ for the models (\ref{eq:S2-13}) are obtained as
\begin{eqnarray}
  \label{eq:S3-9}
  w^1_1 &=& \frac{4}{3} \left(1 + \frac{S^2}{R^2} \frac{kr^2}{(1-kr^2)}\right)^{-1} 
         \left(\frac{\dot{S}}{S} - \frac{\dot{R}}{R}\right) \frac{S^2}{R^2}\frac{kr^2}{(1-kr^2)}\\
\textrm { and }\nonumber\\
w^2_2 &=& w^3_3 = -\frac{2}{3} \left(1 + \frac{S^2}{R^2} \frac{kr^2}{(1-kr^2)}\right)^{-1} 
                 \left(\frac{\dot{S}}{S} -
                   \frac{\dot{R}}{R}\right) \frac{S^2}{R^2} \frac{kr^2}{(1-kr^2)}
\end{eqnarray}
From equation (\ref{eq:S3-5}), the particle number density $n$ for the models (\ref{eq:S2-13}) is obtained as
\begin{equation}
  \label{eq:S3-11}
  n = A(r) R^{-3} \left(1 + \frac{S^2}{R^2} \frac{kr^2}{(1-kr^2)}\right)^{-1/2}
\end{equation}
where $A(r)$ is an arbitrary function of $r$. Since the models (\ref{eq:S2-13}) reduce to $R-W$ models when $S=R$ we obtain $A(r) = (1-kr^2)^{-1/2}$ and therefore,
\begin{equation}
  \label{eq:S3-12}
  n = R^{-3} (1-kr^2)^{-1/2} \left(1+\frac{S^2}{R^2} \frac{kr^2}{(1-kr^2)}\right)^{-1/2}
\end{equation}

The non-vanishing components of the Einstein field equations
\begin{equation}
  \label{eq:S3-13}
  R^i_j - \frac{1}{2} Rg^i_j = -8 \pi T^i_j
\end{equation}
for the line-elements (\ref{eq:S2-13}) and the imperfect fluid distribution (\ref{eq:S3-1}) are obtained as
\begin{equation}
  \label{eq:S3-14}
  \frac{2\ddot{R}}{R} + \frac{\dot{R}^2}{R^2} + \frac{S^2k}{R^4(1-kr^2)} 
  \left(1+ \frac{S^2}{R^2} \frac{kr^2}{(1-kr^2)}\right)^{-1} =
     - 8 \pi [p-\xi\theta-\eta w^1_1],
\end{equation}
\begin{equation}
  \label{eq:S3-15}
    \begin{split}
  \frac{2\ddot{R}}{R} + \frac{\dot{R}^2}{R^2} + \frac{S^2}{R^2} \frac{kr^2}{(1-kr^2)} 
   \left(1+ \frac{S^2}{R^2} \frac{kr^2}{(1-kr^2)}\right)^{-1} 
     \left\{\frac{\ddot{S}}{S} - \frac{\ddot{R}}{R} -
      \frac{\dot{R}^2}{R^2} + \frac{\dot{R} \dot{S}}{R S}\right\} \\
    +\left(1+ \frac{S^2}{R^2} \frac{kr^2}{(1-kr^2)}\right)^{-2}
     \left\{ \frac{S^2k}{r^4(1-kr^2)^2} + \frac{S^2}{R^2} \frac{kr^2}{1-kr^2} 
       (\frac{\dot{S}}{S} - \frac{\dot{R}}{R})^2\right\}\\
      = -8\pi [p-\xi\theta-\eta w^2_2 ],
      \end{split}  
\end{equation}
\begin{equation}
  \label{eq:S3-16}
  \begin{split}
    \frac{3\dot{R}^2}{R^2} + \frac{1}{R^2r^2} + \left(1+ \frac{S^2}{R^2} \frac{kr^2}{(1-kr^2)}\right)^{-1}
      \left\{- \frac{1}{R^2r^2} + \left(\frac{\dot{S}}{S} - \frac{\dot{R}}{R}\right)
        \frac{2S^2}{R^2} \frac{\dot{R}}{R} \frac{kr^2}{(1-kr^2)}\right\}\\
      + \frac{2S^2k}{R^4(1-kr^2)^2} \left(1+ \frac{S^2}{R^2} \frac{kr^2}{(1-kr^2)}\right)^{-2}
           = 8 \pi \rho,
  \end{split}
\end{equation}
and
\begin{equation}
  \label{eq:S3-17}
  \left(1+ \frac{S^2}{R^2} \frac{kr^2}{(1-kr^2)}\right)^{-1}  \left(\frac{\dot{S}}{S} - \frac{\dot{R}}{R}\right)
        \frac{2S^2}{R^2}\frac{kr}{(1-kr^2)} = 8 \pi \chi q_1.
\end{equation}
From equations (\ref{eq:S3-14}) and (\ref{eq:S3-15}) we get
\begin{equation}
  \label{eq:S3-18}
  \begin{split}
    -24 \pi [p-\xi\theta] = \frac{6\ddot{R}}{R} + \frac{3\dot{R}^2}{R^2} + \left(1 +
       \frac{S^2}{R^2} \frac{kr^2}{(1-kr^2)} \right)^{-1}\\
         \left[ \frac{S^2k}{R^4(1-kr^2)} + \frac{2S^2}{R^2} \frac{kr^2}{(1-kr^2)}
           \left\{\frac{\ddot{S}}{S} - \frac{\ddot{R}}{R} - \frac{\dot{R}^2}{R^2} + \frac{\dot{R}}{R}
             \frac{\dot{S}}{S} \right\}\right] \\
         + \left(1 + \frac{S^2}{R^2} \frac{kr^2}{(1-kr^2)} \right)^{-2}
         \left[\frac{2S^2k}{R^4(1-kr^2)^2} + \frac{2S^2}{R^2} \frac{kr^2}{(1-kr^2)}
           \left( \frac{\dot{S}}{S} - \frac{\dot{R}}{R} \right)^2 \right]
  \end{split}
\end{equation}
and
\begin{equation}
  \label{eq:S3-19}
  \begin{split}
    - 16 \pi \eta \left(\frac{\dot{S}}{S} - \frac{\dot{R}}{R}\right) =
       \left(1+\frac{S^2}{R^2} \frac{kr^2}{(1-kr^2)}\right)^{-1}
         \left\{\frac{1}{R^2r^2(1-kr^2)} + \left(\frac{\dot{S}}{S} - \frac{\dot{R}}{R}\right)^2\right\}\\
     + \left\{ \frac{\ddot{S}}{S} - \frac{\ddot{R}}{R} - \frac{\dot{R}^2}{R^2} + 
         \frac{\dot{R}}{R} \frac{\dot{S}}{S}\right\} - \frac{1}{R^2r^2}    
  \end{split}
\end{equation}
From equations (\ref{eq:S3-17}) and (\ref{eq:S3-3}) we get
\begin{equation}
  \label{eq:S3-20}
  \left( 1 + \frac{S^2}{R^2} \frac{kr^2}{(1-kr^2)} \right)^{-1}
     \left( \frac{\dot{S}}{S} - \frac{\dot{R}}{R} \right) \frac{2S^2}{R^2} 
       \frac{kr}{(1-kr^2)} = 8 \pi \chi \frac{\partial T}{\partial r}.
\end{equation}
Integrating this equation we get
\begin{equation}
  \label{eq:S3-21}
  8 \pi \chi T = \left( \frac{\dot{S}}{S} - \frac{\dot{R}}{R} \right)
                    \frac{\left(\frac{S^2}{R^2}\right)}{\left(1-\frac{S^2}{R^2}\right)}
                        \log \left\{ 1 + \left(1 + \frac{S^2}{R^2} 
                            \frac{kr^2}{(1-kr^2)} \right)^{-1} \left(\frac{S^2}{R^2} -1 \right)
                              \right\} + B(t)
\end{equation}
where $B(t)$ is an arbitrary function of time $t$. Since the models (\ref{eq:S2-13}) reduce to R-W models when $S=R$ we get $B(t) = 8\pi \chi T_{R-W}$ and equation (\ref{eq:S3-21}) is now expressed as
\begin{equation}
  \label{eq:S3-22}
  8 \pi \chi (T-T_{R-W}) = \left(\frac{\dot{S}}{S} - \frac{\dot{R}}{R} \right)
          \frac{\left(\frac{S^2}{R^2}\right)}{\left(1-\frac{S^2}{R^2}\right)}
             \log \left\{\frac{\left(\frac{S^2}{R^2}\right)}
               {\left\{1-kr^2(1-\frac{S^2}{R^2})\right\}}\right\}.
\end{equation}

\section{Radiation Dominated Models}

\setcounter{equation}{0}

If the energy-density of the universe is dominated by highly relativistic particles then $p = \frac{1}{3} \rho$ and $\xi = 0$ \cite{Wein72}. From equations (\ref{eq:S3-16}) and (\ref{eq:S3-18}) in this case we get after some simplification
 \begin{equation}
    \label{eq:S4-1}
       \begin{split}
   k^2r^4\left[\frac{S^2}{R^2} \left\{\frac{2\ddot{R}}{R} + \frac{\ddot{S}}{S} + \frac{\dot{R}^2}{R^2}
       + \frac{2\dot{R}}{R}\frac{\dot{S}}{S}\right\} + \frac{R^2}{S^2} \left\{\frac{3\ddot{R}}{R} +
           \frac{3\dot{R}^2}{R^2} \right\}\right. -\\
  \left.\left\{\frac{5\ddot{R}}{R} + \frac{5\dot{R}^2}{R^2} + \frac{\ddot{S}}{S} + \frac{\dot{S}^2}{S^2} \right\}\right] +
   kr^2\left[\left\{\frac{5\ddot{R}}{R} + \frac{5\dot{R}^2}{R^2} + \frac{\ddot{S}}{S} + \frac{\dot{S}^2}{S^2} \right\}\right. -\\
  \left.\frac{2R^2}{S^2} \left\{ \frac{3\ddot{R}}{R} + \frac{3\dot{R}^2}{R^2} \right\} - \frac{k}{R^2} +
     \frac{S^2k}{R^4} \right] +\frac{R^2}{S^2} \left[\frac{3\ddot{R}}{R} + \frac{3\dot{R}^2}{R^2} \right] +
       \frac{3k}{R^2} = 0.
        \end{split} 
  \end{equation}

 This implies
\begin{equation}
   \label{eq:S4-2}
     \begin{split}
      \frac{S^2}{R^2} \left\{\frac{2\ddot{R}}{R} + \frac{\ddot{S}}{S} + 
          \frac{\dot{R}^2}{R^2} + \frac{2\dot{R}}{R}
            \frac{\dot{S}}{S} \right\} + \frac{R^2}{S^2} 
             \left\{\frac{3\ddot{R}}{R} + \frac{3\dot{R}^2}{R^2} \right\}\\
     - \left\{\frac{5\ddot{R}}{R} + \frac{5\dot{R}^2}{R^2} + 
          \frac{\ddot{S}}{S} + \frac{\dot{S}^2}{S^2}\right\} = 0,
   \end{split}   
 \end{equation}
 \begin{equation}
   \label{eq:S4-3}
   \begin{split}
         \left\{\frac{5\ddot{R}}{R} + \frac{5\dot{R}^2}{R^2} + 
            \frac{\ddot{S}}{S} + \frac{\dot{S}^2}{S^2}\right\} 
             - \frac{2R^2}{S^2} \left\{\frac{3\ddot{R}}{R} + \frac{3\dot{R}^2}{R^2} \right\} -
                \frac{k}{R^2} + \frac{S^2k}{R^4} = 0
   \end{split}
 \end{equation}
 and
 \begin{equation}
   \label{eq:S4-4}
   \frac{R^2}{S^2} \left\{ \frac{3\ddot{R}}{R} + \frac{3\dot{R}^2}{R^2} \right\} +
       \frac{3k}{R^2} = 0.
 \end{equation}
 Equation (\ref{eq:S4-4}) gives
 \begin{equation}
   \label{eq:S4-5}
   S^2 = - \frac{R^2}{2k} \frac{d^2}{dt^2} (R^2).
 \end{equation}
 As expected $\frac{d^2}{dt^2}(R^2)$ is positive or negative according as $k = -1$ or $+1$. From equations (\ref{eq:S4-2}) and (\ref{eq:S4-3}) we get
 \begin{equation}
   \label{eq:S4-6}
   \frac{2\ddot{R}}{R} + \frac{\dot{R}^2}{R^2} + \frac{k}{R^2} + \frac{\ddot{S}}{S} +
       \frac{2\dot{S}}{S} \frac{\dot{R}}{R} + \frac{2k}{S^2} = 0.
 \end{equation}
If we take $\frac{\dot{S}}{S} = \frac{\dot{R}}{R}$, that is if we take $S=aR (0 < a \le 1)$ then equations (\ref{eq:S4-5}) and (\ref{eq:S4-6}) imply that $a = 1$, that is the models are R-W models. Making use of equation (\ref{eq:S4-5}) in (\ref{eq:S4-6}) we obtain a differential equation for $R^2$ as
 \begin{equation}
   \label{eq:S4-7}
   \begin{split}
     \frac{d^4}{dt^4} (R^2) - \frac{1}{2} \left\{\frac{d^3}{dt^3} (R^2) \right\}^2
        \left\{\frac{d^2}{dt^2} (R^2)\right\}^{-1} + \frac{2}{R^2} \frac{d}{dt}(R^2) 
          \frac{d^3}{dt^3}(R^2)\\
      + \frac{3}{R^2} \left\{ \frac{d^2}{dt^2} (R^2)\right\}^2 + \frac{2k}{R^2}
          \left\{\frac{d^2}{dt^2} (R^2) \right\} - \frac{8k^2}{R^2} = 0.
   \end{split}
 \end{equation}
 Equation (\ref{eq:S3-6}) in this case takes the form
 \begin{equation}
   \label{eq:S4-8}
   d(K\sigma) = \frac{4\rho}{3T} d\left(\frac{1}{n}\right) + \frac{1}{nT} d\rho.
 \end{equation}
 This implies
 \begin{equation}
   \label{eq:S4-9}
   T = 3n^{1/3}
 \end{equation}
 and
 \begin{equation}
   \label{eq:S4-10}
   K\sigma = \frac{1}{3} \rho n^{-4/3}.
 \end{equation}
 In view of equation (\ref{eq:S3-12}) we get equation (\ref{eq:S4-9})
 \begin{equation}
   \label{eq:S4-11}
   T = 3\left[R\left\{(1-kr^2)\left(1+\frac{S^2}{R^2} \frac{kr^2}{(1-kr^2)}
         \right)\right\}^{1/6}\right]^{-1}
 \end{equation}
 where $S$ is given in terms of $R$ from equation (\ref{eq:S4-5}) and $R$ is
 given by the differential equation (\ref{eq:S4-7}). Making use of equation 
 (\ref{eq:S4-11}) in (\ref{eq:S3-22}) we obtain
\begin{equation}
   \label{eq:S4-12}
     \begin{split}
  {8 \pi \chi \left\{ 3R^{-1} (1-kr^2)^{-1/6}
       \left(1+\frac{S^2}{R^2} \frac{kr^2}{(1-kr^2)}\right)^{-1/6}
         -3(R_{rad})^{-1}\right\}}\\
  = \left(\frac{\dot{S}}{S} - \frac{\dot{R}}{R}\right) \frac{\left(\frac{S^2}{R^2}\right)}
         {\left\{1-\frac{S^2}{R^2}\right\}} \log \left\{\frac{\frac{S^2}{R^2}}
           {\left\{1-kr^2\left(1-\frac{S^2}{R^2}\right)\right\}}\right\}
       \end{split}
\end{equation}
 where $R_{rad}$ is the scale factor of the R-W models in radiation dominated
 era. With the help of equations(\ref{eq:S3-12}) and (\ref{eq:S3-16}) we can
 write equation (\ref{eq:S4-10}) as
 \begin{equation}
   \label{eq:S4-13}
   \begin{split}
      K \sigma = \frac{1}{24 \pi} \left[\frac{3\dot{R}^2}{R^2} + \frac{1}{R^2r^2} +
         \left(1 + \frac{S^2}{R^2} \frac{kr^2}{(1-kr^2)}\right)^{-1}\right.\\
   \left.\left\{-\frac{1}{R^2r^2} + 
       \left(\frac{\dot{S}}{S} - \frac{\dot{R}}{R}\right)
        \frac{2S^2}{R^2} \frac{\dot{R}}{R} \frac{kr^2}{(1-kr^2)}  \right\}+
          \frac{2S^2k}{R^4(1-kr^2)^2} \left(1+ \frac{S^2}{R^2} \frac{kr^2}{(1-kr^2)}
            \right)^{-2}\right]\\
    \left[R^4(1-kr^2)^{2/3} \left(1 + \frac{S^2}{R^2} \frac{kr^2}{(1-kr^2)}
         \right)^{2/3}\right]
   \end{split}
 \end{equation}
 where $S$ is given in terms of $R$ from equation (\ref{eq:S4-5}) and $R$ is given 
 by the differential equation (\ref{eq:S4-7}).

\section{Matter Dominated Models}

\setcounter{equation}{0}

If the energy density of the universe is dominated by the non-relativistic
matter with negligible pressure and bulk viscosity then equation (\ref{eq:S3-18})
gives after some simplification

\begin{equation}
   \label{eq:S5-1}
   \begin{split}
       k^2 r^4  \left[ \frac{S^4}{R^4} \left\{\frac{4\ddot{R}}{R} + \frac{2\ddot{S}}{S} +
          \frac{\dot{R}^2}{R^2} + \frac{2\dot{R}}{R}\frac{\dot{S}}{S}\right\}  - \frac{S^2}{R^2}
            \left\{\frac{10\ddot{R}}{R} + \frac{2\ddot{S}}{S} + \frac{6\dot{R}^2}{R^2} +\right.\right.\\
        \left.\left.\frac{2\dot{S}^2}{S^2} - \frac{2\dot{R}}{R}\frac{\dot{S}}{S} \right\} +
          \left\{\frac{6\ddot{R}}{R} + \frac{3\dot{R}^2}{R^2} \right\}\right]
           + kr^2 \left[\frac{S^2}{R^2} \left\{\frac{10\ddot{R}}{R} + \frac{2\ddot{S}}{S} +
           \frac{6\dot{R}^2}{R^2} + \frac{2\dot{S}^2}{S^2} - \frac{2\dot{R}}{R} 
             \frac{\dot{S}}{S}\right\}\right.\\
        \left. -2\left\{\frac{6\ddot{R}}{R} + \frac{3\dot{R}^2}{R^2} \right\} +
           \left\{\frac{kS^4}{R^6} - \frac{kS^2}{R^4} \right\}\right] +
               \frac{6\ddot{R}}{R} + \frac{3\dot{R}^2}{R^2} + \frac{3kS^2}{R^4} = 0.
   \end{split}
 \end{equation}
  This implies
 \begin{equation}
   \label{eq:S5-2}
   \begin{split}
       \frac{S^4}{R^4} \left\{\frac{4\ddot{R}}{R} + \frac{2\ddot{S}}{S} +
           \frac{\dot{R}^2}{R^2} + \frac{2\dot{R}}{R}\frac{\dot{S}}{S} \right\}-
              \frac{S^2}{R^2} \left\{\frac{10\ddot{R}}{R} + \frac{2\ddot{S}}{S} +\right.\\
       \left.\frac{6\dot{R}^2}{R^2} + \frac{2\dot{S}^2}{S^2} - 
           \frac{2\dot{R}}{R} \frac{\dot{S}}{S}\right\}
            + \left\{\frac{6\ddot{R}}{R} + \frac{3\dot{R}^2}{R^2} \right\} = 0,
   \end{split}
 \end{equation}

\begin{equation}
  \label{eq:S5-3}
    \begin{split}
   \frac{S^2}{R^2} \left\{\frac{10\ddot{R}}{R} + \frac{2\ddot{S}}{S} + \frac{6\dot{R}^2}{R^2}
         + \frac{2\dot{S}^2}{S^2} - \frac{2\dot{R}}{R}\frac{\dot{S}}{S} \right\} - 2
             \left\{\frac{6\ddot{R}}{R} + \frac{3\dot{R}^2}{R^2} \right\}\\
       + \left\{ \frac{kS^4}{R^6} - \frac{kS^2}{R^4} \right\} = 0
    \end{split}
\end{equation}
 and
\begin{equation}
  \label{eq:S5-4}
  \frac{6\ddot{R}}{R} + \frac{3\dot{R}^2}{R^2} + \frac{3kS^2}{R^4} = 0.
\end{equation}
From equation (\ref{eq:S5-4}) we get
\begin{equation}
  \label{eq:S5-5}
  S^2 = - \frac{R^2}{k} (2\ddot{R} R + \dot{R}^2)
\end{equation}
This shows that $(2 \ddot{R} R + \dot{R}^2)$ is positive or negative according
as $k$ is $-1$ or $+1$. From equations (\ref{eq:S5-2}) and (\ref{eq:S5-3})
we get
\begin{equation}
  \label{eq:S5-6}
  \left(\frac{4\ddot{R}}{R} + \frac{\dot{R}^2}{R^2} + \frac{k}{R^2} \right) +
      \frac{2\ddot{S}}{S} + \frac{2\dot{R}}{R} \frac{\dot{S}}{S} + \frac{2k}{S^2} = 0
\end{equation}
If we take $\frac{\dot{S}}{S} = \frac{\dot{R}}{R}$ then in this case also equations
(\ref{eq:S5-5}) and (\ref{eq:S5-6}) imply that the models are $R-W$ models. 
If we substitute for $S$ from equation (\ref{eq:S5-5}) in (\ref{eq:S5-6}) 
we again get a differential equation of fourth order for $R$.

Equation (\ref{eq:S3-6}) in this case leads to
\begin{equation}
  \label{eq:S5-7}
  T = constant \, (say \, M)
\end{equation}
and
\begin{equation}
  \label{eq:S5-8}
  K\sigma = \frac{\rho}{Mn}.
\end{equation}
Equation (\ref{eq:S3-22}) in this case gives
\begin{equation}
  \label{eq:S5-9}
  8 \pi \chi (M - \bar{T}_{R-W}) = \left(\frac{\dot{S}}{S} - \frac{\dot{R}}{R} \right)
       \frac{\left(\frac{S^2}{R^2}\right)}{\left\{1-\frac{S^2}{R^2}\right\}}
          \log \left\{\frac{\left(\frac{S^2}{R^2}\right)}
             {\left\{1-kr^2(1-\frac{S^2}{R^2})\right\}}\right\}
\end{equation}
where $\bar{T}_{R-W}$ is the constant temperature of the R-W models in the matter
dominated era. Making use of equations (\ref{eq:S3-12}) and (\ref{eq:S3-16}) 
in equation (\ref{eq:S5-8}) we get
\begin{equation}
  \label{eq:S5-10}
  \begin{split}
    K\sigma = \frac{1}{8 \pi M} \left[ \frac{3\dot{R}^2}{R^2} + \frac{1}{R^2r^2} +
                \left(1 + \frac{S^2}{R^2} \frac{kr^2}{(1-kr^2)}\right)^{-1}.
                  \left\{- \frac{1}{R^2r^2} +\right.\right.\\
    \left.\left.\left(\frac{\dot{S}}{S} - \frac{\dot{R}}{R}\right) \frac{2S^2}{R^2} \frac{\dot{R}}{R}
         \frac{kr^2}{(1-kr^2)}\right\} + \frac{2S^2k}{R^4(1-kr^2)^2}
           \left(1+\frac{S^2}{R^2} \frac{kr^2}{(-kr^2)}\right)^{-2} \right]\\
       \left[R^3(1-kr^2)^{1/2} \left(1+\frac{S^2}{R^2} \frac{kr^2}{(1-kr^2)}\right)^{1/2}\right].
  \end{split}
\end{equation}
In the above two equations (\ref{eq:S5-9}) and (\ref{eq:S5-10}), $S$ is given in terms
of $R$ from equation (\ref{eq:S5-5}) and $R$ is given by the differential equation
obtained from (\ref{eq:S5-6}) after substituting for $S$ from (\ref{eq:S5-5}).

\section{The Hubble and Deacceleration parameters}

\setcounter{equation}{0}

A representative length $l$ which represents the volume behaviour of the 
fluid is defined by the equation
\begin{equation}
  \label{eq:S6-1}
\frac{l^*}{l} \equiv \frac{l_{;i}u^i}{l} = \frac{1}{3}\theta=\frac{\dot l}{l}.
\end{equation}
For the models (\ref{eq:S2-13}) $l$ is obtained as
\begin{equation}
  \label{eq:S6-2}
  l = C(r) R\left(1+\frac{S^2}{R^2} \frac{kr^2}{(1-kr^2)}\right)^{1/6}
\end{equation}
where $C(r)$ is an arbitrary function of $r$. Since the models (\ref{eq:S2-13}) reduce to R-W
models when $S=R$ we get $C(r)=(1-kr^2)^{1/6}$ and therefore,
\begin{equation}
  \label{eq:S6-3}
  l = R(1-kr^2)^{1/6} \left(1+\frac{S^2}{R^2} \frac{kr^2}{(1-kr^2)}\right)^{1/6}.
\end{equation}
The Hubble parameter $H$ is defined by the equation
\begin{equation}
  \label{eq:S6-4}
  H = \frac{l^*}{l}.
\end{equation}
For the models (\ref{eq:S2-13}), $H$ is given by
\begin{equation}
  \label{eq:S6-5}
  H = \frac{\dot{R}}{R} + \frac{1}{3} \left( 1+ \frac{S^2}{R^2} 
       \frac{kr^2}{(1-kr^2)}\right)^{-1} \left( \frac{\dot{S}}{S} -
          \frac{\dot{R}}{R}\right)\frac{S^2}{R^2}\frac{kr^2}{(1-kr^2)}.
\end{equation}
The deacceleration parameter $q$ is defined by the equation
\begin{equation}
  \label{eq:S6-6}
  q = - \left( \frac{l^{**}}{l}\right)\left(\frac{l}{l^*}\right)^2.
\end{equation}
For the models (\ref{eq:S2-13}) $q$ is obtained as
\begin{equation}
  \label{eq:S6-7}
  \begin{split}
    q = -1-\left\{\frac{\dot{R}}{R} + \frac{1}{3} \frac{S^2}{R^2}
         \frac{kr^2}{(1-kr^2)} \left( 1 + \frac{S^2}{R^2}
            \frac{kr^2}{(1-kr^2)}\right)^{-1} \left( \frac{\dot{S}}{S}
              - \frac{\dot{R}}{R}\right)\right\}^{-2}.\\
  \left[\left( \frac{\ddot{R}}{R} - \frac{\dot{R}^2}{R^2}\right) + \frac{1}{3}
      \frac{S^2}{R^2} \frac{kr^2}{(1-kr^2)} \left( 1 + 
        \frac{S^2}{R^2} \frac{kr^2}{(1-kr^2)}\right)^{-1}\right.\\
   \left.\left\{\left(\frac{\ddot{S}}{S} - \frac{\ddot{R}}{R} - \frac{\dot{S}^2}{S^2} +
        \frac{\dot{R}^2}{R^2}\right) + 2 \left(\frac{\dot{S}}{S} -
           \frac{\dot{R}}{R}\right)^2 \left( 1 + \frac{S^2}{R^2}
              \frac{kr^2}{(1-kr^2)}\right)^{-1} \right\}\right].
  \end{split}
\end{equation}
From equations (\ref{eq:S6-1}), (\ref{eq:S6-4}) and (\ref{eq:S6-6})
we find that
\begin{equation}
  \label{eq:S6-8}
  q = -1 + \frac{d}{dt} (H^{-1}).
\end{equation}
With the help of equations (\ref{eq:S6-3}), (\ref{eq:S6-5}) and (\ref{eq:S6-7})
the expression for matter-energy density given by equation (\ref{eq:S3-16}) and the expression for the isotropic pressure given by equation (\ref{eq:S3-18}) can be expressed as
\begin{equation}
  \label{eq:S6-9}
  8 \pi \rho = 3\left(\frac{\dot{R}^2}{R^2}\right)\left(2H \left(
       \frac{\dot{R}}{R}\right)^{-1} -1\right)+\left\{1 +
          \frac{2R^6}{l^6}\right\}\left\{\frac{l^6-R^6(1-kr^2)}
             {l^6R^2r^2}\right\}
\end{equation}
or
\begin{equation}
  \label{eq:S6-10}
  8 \pi \rho = 3(H_{R-W})^2 \left\{2H (H_{R-W})^{-1}-1\right\}+
      \left\{1 + \frac{2R^6}{l^6}\right\}\left\{\frac{l^6-R^6
         (1-kr^2)}{l^6R^2r^2}\right\}
\end{equation}
and
\begin{equation}
  \label{eq:S6-11}
  \begin{split}
    8 \pi (p -\xi\theta) = 2 (1+q)H^2 - \frac{3\dot{R}^2}{R^2} -
       \frac{\left\{l^6-R^6(1-kr^2)\right\}}{l^6R^2r^2}\left(
         1 + \frac{2R^6}{l^6}\right) -\\
    \frac{6(H-\frac{\dot{R}}{R})^2 l^6}{\left\{l^6 - R^6 (1-kr^2)\right\}}
       \left\{1 - \frac{2R^6(1-kr^2)}{l^6} \right\} - 6\left(H-\frac{\dot{R}}{R}\right)
          \frac{\dot{R}}{R}
  \end{split}
\end{equation}
or
\begin{equation}
  \label{eq:S6-12}
  \begin{split}
    8 \pi (p - \xi \theta) = 2 (1 + q) H^2 -
       3(H_{R-W})^2 - \frac{\{l^6 - R^6 (1-kr^2)\}}{l^6R^2r^2}
         \left(1 + \frac{2R^6}{l^6}\right)- \\
    \frac{6(H - H_{R-W})^2 l^6}{\{l^6-R^6(1-kr^2)\}} 
    \left\{1 - \frac{2R^6(1-kr^2)}{l^6}\right\}- 6(H-H_{R-W})
          H_{R-W}.
  \end{split}
\end{equation}


\thebibliography{9}
\bibitem{Nar93}
Narlikar, J.V. 1993,
\textit{Introduction to Cosmology},
Cambridge University Press.

\bibitem{Wein72}
Weinberg, S. 1972,
\textit{Gravitation and Cosmology : Principles and Applications of the General Theory of Relativity},
John Wiley \& Sons, New York.
\end{document}